\documentclass[11pt]{iopart}

\usepackage{epsfig,graphicx,latexsym,iopams}
\usepackage{color}

\DeclareGraphicsExtensions{.eps}

\newcommand{\figu}{Fig.~}

\newcommand{\sect}{Section~}

\newcommand{\sis}{$\sigma$}

\newcommand{\mbh}{$M_{\rm BH}$}
\newcommand{\mbhe}{M_{\rm BH}}
\newcommand{\mstar}{$M_{\rm star}$}
\newcommand{\mstare}{M_{\rm star}}
\newcommand{\msunhe}{\rm{M_{\odot}\, h^{-1}}}

\newcommand{\Vmax}{$V_{\rm max}$}

\newcommand{\msune}{M_{\odot}}

\newcommand{\mh}{$M_H$}
\newcommand{\mhe}{M_H}

\def\ls{\lower 2pt \hbox{$\;\scriptscriptstyle \buildrel<\over\sim\;$}}
\def\gs{\lower 2pt \hbox{$\;\scriptscriptstyle \buildrel>\over\sim\;$}}

\begin{document}

\title[Black Hole Demography]{Black Hole Demography: From scaling relations to models}

\author{Francesco Shankar}

\address{GEPI, Observatoire de Paris, CNRS, Univ. Paris Diderot, 5 Place Jules Janssen, F-92195 Meudon, France}
\ead{francesco.shankar@obspm.fr}
\begin{abstract}
In this contributed paper I review our current
knowledge of the local Black Hole (BH) scaling relations,
and their impact on the determination of the local BH mass function.
I particularly emphasize the remaining systematic uncertainties
impinging upon a secure determination of the BH mass function
and how progress can be made.
I then review and discuss the evidence for
a different time evolution for separate BH-galaxy scaling relations,
and how these independent empirical evidences can be reconciled with the
overall evolution of the structural properties of the host galaxies.
I conclude discussing BH demography in the context of semi-empirical continuity accretion models,
as well as more complex evolutionary models, emphasizing the general constraints
we can set on them.
\end{abstract}

\maketitle

\label{firstpage}


\section{Introduction}
\label{sec|intro}

Supermassive black holes (BHs) reside at the center of most, if not all,
massive galaxies.  The masses of BHs are tightly
correlated with properties of their galactic hosts, especially
the velocity dispersions and masses of their stellar bulges
\cite{Ferrarese00,Gebhardt00,HaringRix,Gultekin09}.

However, while it is now clear that relations between BHs and their
hosts exist, it is still a matter of hot debate in the recent Literature what
their slopes, intrinsic scatters, and evolution with redshift, truly are.
This is partly due to the still limited working samples
of independent BH mass measurements available,
as detailed kinematical studies are limited to mostly nearby galaxies.
Nevertheless, the number of secure BH measurements
has increased over the years, and advances have been achieved.

Given the interesting progress made in the recent Literature \cite{BongiornoReview},
parallelling the strong interest of the astronomical community
in the broad topic of BH demography, it is valuable to summarize some of the
key recent results and try to insert them in a coherent context.
The aim of this contributed paper is thus to
provide the reader with a short guide through a variety of different empirical
and theoretical recent results, revisited with the intent of
connecting them within a plausible, broad physical framework.

\section{News about the local Black Hole scaling relations}
\label{sec|scalingrelations}

In the local universe, BH masses \mbh, correlate with several global properties of the host galaxies.
One of the most studied one is between BH mass and host stellar (bulge or total) luminosity/mass \cite{Magorrian98}.
The usual trend usually quoted in the Literature
is a linear correlation between BH mass and host bulge luminosity,
with a slope close to unity and a normalization of about $1-2\times 10^{-3}$ \cite{Marconi03,HaringRix}.
Recent studies performed in the NIR/IR bands, less affected by dust extinction and
more sensitive to total stellar mass, tend to confirm this trend with a slope
about $\sim 0.9$ \cite{Sani11,Vika12}, or slightly shallower $\sim 0.8$ (L\"{a}sker et al. 2013, submitted).
The scatter of this relation, initially claimed to be around 0.5 dex in bluer bands \cite{Kormendy95},
has been somewhat reduced to $\sim 0.3-0.4$ dex in the latest calibrations \cite{Sani11,Vika12,McConn13}, making
it closer to the scatter in the \mbh-\sis\ relation.

Possibly one of the newest proposals in this respect has been put forward by \cite{Graham12},
who claimed for a net break in the BH-bulge mass relation, dependent on the host galaxy profile.
Core S\'{e}rsic galaxies, mostly dominating above $\mbhe \gtrsim 2\times 10^8\, \msune$, will continue following
a linear relation between BH mass and bulge mass, while S\'{e}rsic galaxies at lower masses
tend to follow a quadratic relation with host bulge mass. This claim was further
quantified by \cite{Scott13}.

On more general grounds, it has been several times emphasized in the very last
years that, especially at the low BH-mass end, BH scaling relations
tend to exhibit larger scatters, where the hosts preferentially become later-type systems.
\cite{KormendyHo} have reviewed this topic claiming that BHs
BHs correlate differently with different galaxy components.
In particular, they stress that any correlation with
disc-grown pseudobulges or halo dark matter are very weak, implying no close co-evolution.
In fact, as pointed out by \cite{ShankarPseudo},
at face value the possible large scatter in local BH-bulge mass relation induced when
including all measurements with no restriction on galaxy type, is hard to reproduce by models
in which the fuelling of BHs closely follows the growth of their host bulges,
such as in late bar-instability modes (see \sect\ref{sec|Models}).

The correlation with velocity dispersion \sis\ continues to hold the record as the tightest correlation.
\cite{Beifiori12} have recently re-analyzed all the correlations between BH mass
and host galaxy property, including S\'{e}rsic index, circular velocity, and galaxy dynamical
and effective masses. They confirm there is no evidence for a tighter correlation
than the one between \mbh\ and \sis, at least for the so-called ``classical'' bulges,
and that the correlation with large scale quantities such as the circular velocity
is weaker. They then tested the need for a third parameter in
the BH scaling relations, confirming that the Fundamental
Plane of BHs \cite{Hop07FP} is mainly driven by \sis, with a small tilt due to
the effective radius.

\cite{Beifiori12} also claimed a poor correlation between BH mass and S\'{e}rsic index, at variance
with previous findings. \cite{Savorgnan13}
revisited the issue of the relation between
BH mass and the central light concentration of the surrounding bulge,
quantified by the S\'ersic index $n$. They claimed
that a clear correlation exists, although with
significant scatter. They then discuss how this relation is consistent
with what would be derived by combining the $\mbhe-L_{\rm sph}$ and
$L_{\rm sph}-n$ relations, and conclude on how, for the same central light concentration,
the correlation with BH mass could change with galaxy profile.

\section{Probing the Local Black Hole Mass Function}
\label{sec|BHMF}

Improved measurements of the local BH scaling relations are clearly
fundamental to further advance in our true knowledge of the co-evolution
between BHs and galaxies. Moreover, more secure BH scaling relations
can potentially set interesting constraints on the number density
of BHs as a function of mass and time, providing in turn
useful terms of comparisons for models. For instance, some groups
place a stronger emphasis on BH growth through gas accretion \cite{Granato04,Lapi06},
while others claim that growth by mergers plays an important role
especially for the most massive BHs \cite{Malbon07}, thus possibly
impacting the shape of the high mass end BH mass function \cite{Shankar13acc}.
Similarly, growth histories characterized by distinctive time and/or mass dependent
BH accretion rates, can easily yield at $z=0$ quite different BH number densities
at low to intermediate mass scales \cite{SWM}.

The advent of large, well studied, galaxy surveys
such as the Sloan Digital Sky Survey (SDSS),
has allowed through time an improved understanding of the local
demography of galaxies in terms of luminosity, stellar mass,
size, and velocity dispersion \cite{Bernardi10,Bernardi13}.
Coupling this information with the above mentioned BH-galaxy scaling relations,
then allows to calibrate the total mass distribution of BHs
\cite{ShankarReview,KellyMerloni}.
The most basic procedure for calculating the BH mass function has
been to assume that all galaxies host one BH.
The local BH mass function is then derived by converting
the galaxy distribution $\Phi(y)$, expressed as a function of a
given measured variable $y$ (e.g., the stellar velocity dispersion
or bulge luminosity/stellar mass), into a BH mass function by assuming a corresponding
empirical \mbh-$y$ relation. Specifically, the BH mass function is
computed via the equation \cite{Salucci99}
\begin{equation}
\Phi(\mbhe)=\int \Phi(y) \frac{1}{\sqrt{2\pi
\eta^2}}\exp\left[-\frac{\left(\mbhe-\left[a+by \right]
\right)^2}{2\eta^2}\right]dy\nonumber
    \label{eq|BHMFscatter}
\end{equation}
which also accounts for the scatter $\eta$ in the \mbh-$y$ relation.

Alternatively, one could directly use a complete galaxy catalog,
assign a BH to each galaxy via the chosen empirical scaling relation,
and then determine BH mass functions directly using
the \Vmax\ weight appropriate for each galaxy
\cite{Graham07BHMF,Vika09}.
In a statistical sense, the two methods should provide equivalent results.

\subsection{The Normalization issue}
\label{NormalizationBHMF}

Obtaining good estimates of the BH
mass function is not a trivial task,
and in fact different approaches and assumptions
may lead to different answers.
For example, several groups
point to significant discrepancies both in the
shape and normalization of the BH mass function when
adopting different scaling relations \cite{Tundo07,Graham07,Vika09}.

To highlight this point, in this section I show two
estimates of the local BH mass function derived from
up-to-date scaling relations and galaxy functions.
In the specific, I use the early-type luminosity and velocity dispersion functions by
\cite{Bernardi10}, converted into BH mass functions using the
BH mass-$r$ band luminosity, \mbh-$L_r$, and BH mass-velocity dispersion, \mbh-\sis,
relations of early-type galaxies by \cite{Graham07} and \cite{McConn13}, respectively.
The BH mass function $\pm 1\, \sigma$ error bars are computed via Monte Carlo simulations \cite{Marconi04}.
In practice, I collect the results of 1000 realizations
in which I allow to vary within the measured uncertainties
the parameters of the \mbh-$y$ relation, its scatter $\eta$,
and the Poisson errors associated to $\Phi(y)$.
For each bin of BH mass, I then compute the median and $1\, \sigma$ errors
associated to the distribution of $\log \Phi(\mbhe)$.

\begin{figure*}
    \center{\includegraphics[width=15truecm]{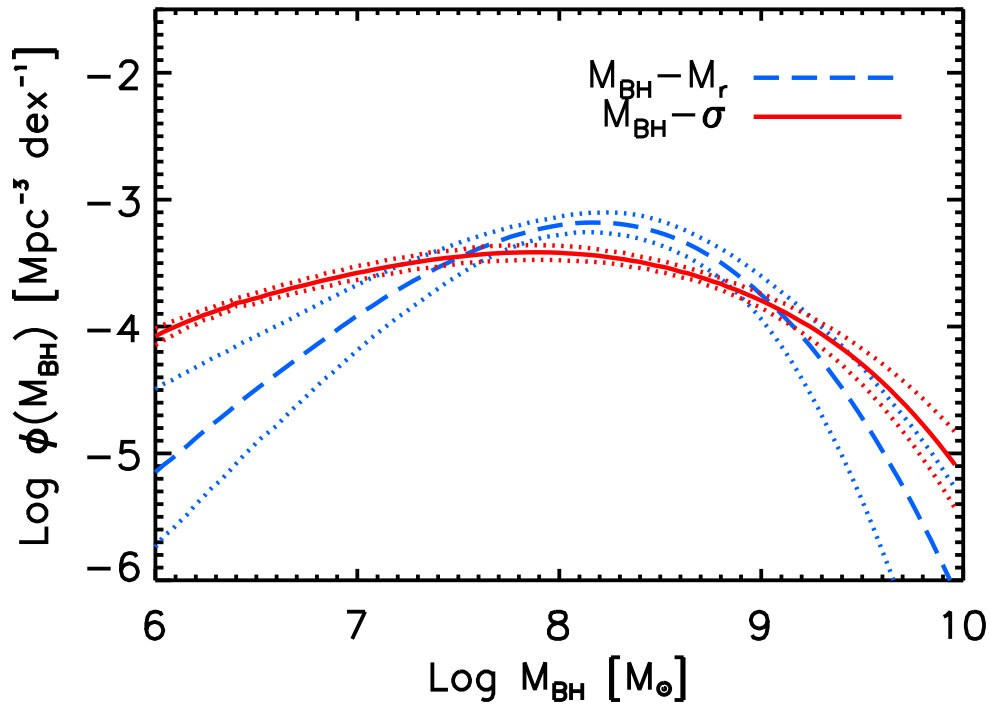}
    \caption{Comparison between the local BH mass functions
    of ellipticals derived from the stellar mass (\emph{long-dashed}, \emph{blue} line)
    and velocity dispersion (\emph{solid}, \emph{red} line)
    function by \cite{Bernardi10}, and \mbh-$L_r$ and \mbh-\sis\ relations by \cite{Graham07} and
    \cite{McConn13}, respectively. \emph{Dotted} lines refer to the $1\, \sigma$ uncertanties computed via Monte Carlo
    simulations.}}
\label{fig|LocalBHMFinEs}
\end{figure*}

The result, shown in \figu\ref{fig|LocalBHMFinEs}, proves the importance of well calibrating
the local scaling relations in order to estimate a more reliable BH mass function.
The example in \figu\ref{fig|LocalBHMFinEs} is particularly meaningful in two respects.
First, the statistics of galaxies in stellar mass and velocity space is homogeneous, being derived from the
same exact SDSS subsample. Second, the analysis is restricted to a subsample dominated by ellipticals,
thus no bulge correction has been included in the calculation.
We only convert from the SDSS $r$-band to the $R$ band in \cite{Graham07}
adopting an average colour correction of $r-R=0.2$ \cite{Fukugita95}.
This implies that the differences in the final BH mass functions,
highly significant in some bins ($\gtrsim 2\, \sigma$), are mainly driven by our choice of scaling relations.
Similar, if not larger, discrepancies would have been obtained by using, for instance, stellar
masses in place of galaxy luminosities.
An analogous BH mass function results in fact
by using, for example, the \mbh-\mstar\ by \cite{Sani11},
and assuming an average correction of 0.25 dex,
typical of massive galaxies \cite{Bernardi10},
to convert from dynamical masses to a Chabrier initial mass function \cite{Chabrier03}.

\subsection{The Low Mass End}
\label{subsec|LowMassEnd}

As anticipated in \sect\ref{sec|scalingrelations},
possibly not all galaxies may follow the same tight
scaling relations in the local universe.
In particular, a number of independent groups seem to find that especially for
galaxies hosting the so-called ``pseudobulges'',
there is a near absence of correlation between BH mass and
velocity dispersion/bulge stellar mass of the host, with BHs
being often significantly less massive than what expected by classical
scaling relations \cite{Hu09,Greene10,Kormendy11,Sani11,Beifiori12}.
\cite{Hu09} also claimed that pseudobulges follow
scaling relations a factor of $\sim 3$, while \cite{Kormendy11}
do not find signs for any evident correlation for BHs residing in pseudobulges.
\cite{McConn13} find that late-type galaxies follow a \mbh-\sis\ relation
with similar slope and scatter as for early-type galaxies, but with a lower
normalization lower by a factor of $\sim 3$.
Also the break in the \mbh-\mstar\ relation advocated by \cite{Scott13}
could at some level be reconciled with the findings by \cite{McConn13,Hu09}.

Moreover, it has been suggested that the
vast majority of low mass galaxies may host pseudobulges \cite{Fisher11}.
Under this working assumption,
\cite{ShankarPseudo} showed that assuming that all
Sb galaxies host pseudobulges, the BH mass function could
lead to a significantly reduced number density at low masses.
Properly studying the impact of pseudobulges
on the local BH mass function is far from trivial.
One possibility could be to
start from the Hubble-dependent stellar mass
function by, e.g., \cite{Bernardi10}, and then
to correct it adopting the stellar mass-dependent fractions by \cite{Fisher11}.
However, the latter approach suffers from the significant uncertainties
associated to the method of using stellar masses to infer BH masses
\cite{SWM,Sani11}.

\begin{figure*}
    \center{\includegraphics[width=15truecm]{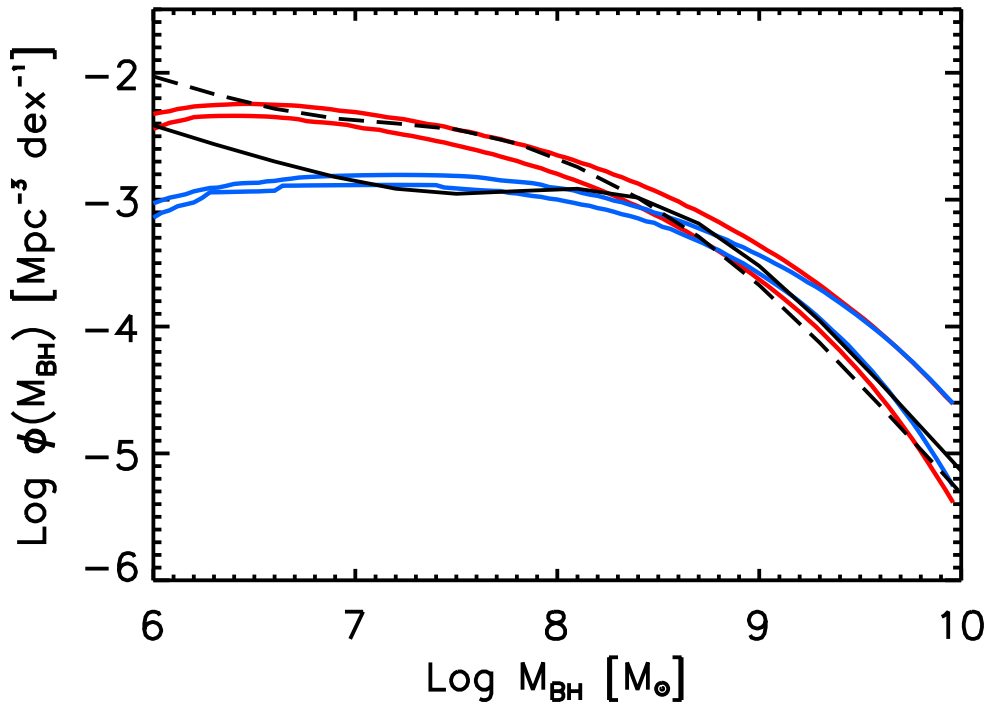}
    \caption{The \emph{red}, \emph{solid} lines show the $1\, \sigma$ uncertainty
    region for the local BH mass function derived on the assumption
    that all local galaxies follow
    the early-type \mbh-\sis\ relation by \cite{McConn13}. The \emph{blue}, \emph{solid} lines
    mark the $1\, \sigma$ uncertainty region inferred assuming the BH mass in Sa galaxies
    is negligible. The \emph{long-dashed} and \emph{solid} lines show the $z=0$ BH mass function
    derived from the \cite{Shankar13acc} continuity equation models assuming a characteristic
    Eddington ratio constant and decreasing with cosmological time (\emph{long-dashed} and \emph{solid} lines, respectively).}}
\label{fig|BHmassFunctionPseudo}
\end{figure*}

Following the strategy pursued by \cite{ShankarPseudo}, to highlight
the possible impact of pseudobulges,
we present a modified BH mass function in \figu\ref{fig|BHmassFunctionPseudo}.
The red, solid lines bracket the $1\, \sigma$ uncertainty
region for the local BH mass function derived on the assumption
that all local galaxies follow
the early-type \mbh-\sis\ relation by \cite{McConn13}. The blue, solid lines
mark instead the $1\, \sigma$ uncertainty region inferred by assuming that BH masses in Sa galaxies
are negligible. As discussed by \cite{Greene08,GadottiKauffmann},
and preliminary quantified by \cite{ShankarPseudo},
allowing for a significant fraction of BHs to be hosted in pseudobulges,
can indeed have a major impact on the
local BH mass function, as seen in \figu\ref{fig|BHmassFunctionPseudo},
decreasing the number density by a factor of
$\sim 2$ around $\mbhe \sim 10^8 \msune$, up to nearly an order of magnitude for $\mbhe \lesssim 10^6 \msune$.
We thus conclude that determining a secure estimate of the local BH mass
function requires detailed knowledge of the role of pseudobulges,
nuclear star clusters \cite{Cote06}, breaks or Hubble-dependent variations
in the BH scaling relations, etc...
What has been calibrated in this work (and most of previous ones)
may safely be considered as an actual \emph{upper limit}
to the true function describing the demography of local BHs.

\section{The Evolution of Black Hole-Host scaling relations}
\label{sec|EvolScalingRelations}

Beyond the local universe, a variety of statistical studies on the
BH accretion history (discussed in \sect\ref{subsec|Accretion}),
support the view that the redshift evolution of median BH accretion and
star formation rate track each other
\cite{Marconi04,Merloni04,SWM},
consistent with the general idea that massive BHs and
their host galaxies may indeed co-evolve at some level.
However, the latter are hints derived from integrated quantities, and are affected
by systematic uncertantities (although recent studies carried out by \cite{Mulla12,Shankar13acc}
continue to support a close link even in the ratio between average BH accretion rate and
host galaxy star formation rate).

One way to test BH-galaxy co-evolution,
and at the same time to begin to explore the evolution of the BH mass function,
is by direct and indirect measurements of the cosmic evolution of the scaling relations.
A variety of studies have tried in the last decade or so to infer the degree
of evolution in the scaling relations of BHs, in particular the BH-stellar mass and BH-velocity dispersion relations. A positive/negative evolution in these scaling relations
could physically imply that, on average, BHs grow faster/slower than their host galaxies,
thus suggesting some non-parallel evolution between the two systems.

Among the first, \cite{McLure06} measured
the BH-to-host galaxy mass ratio in a sample of radio-loud Active Galactic Nuclei (AGN)
in the redshift range $0<z<2$ supporting a strong evolution close to $\mbhe/\mstare \propto (1+z)^2$.
More recently, their work was confirmed by different groups working with larger and different types of samples
\cite{DeCarli10,Bennert11}.

As pointed by \cite{Lauer07bias,VolonteriStark}, however, an increasing scatter at higher redshifts
in the scaling relations could clearly mimic/increase the effect of evolution in the
normalization in flux limited samples.
At intermediate redshifts $0.4 \lesssim z \lesssim 2.5$, the constraints from quasar clustering
under the assumption of a monotonic mean relation between quasar luminosity and host halo mass,
support independent evidence of a relatively large lognormal scatter \cite{Shankar10shen}.
Based on a large and multiwavelength sample extracted from zCOSMOS,
\cite{Merloni10} claimed instead a much milder, but still significant,
evolution of the type $\mbhe/\mstare \propto (1+z)^{0.68\pm0.12}$,
along with a sign for a possible increase in the scatter.

However, any strong evolution in scatter should break down or somewhat stabilize at redshifts
higher than $z\sim 3$, at least for the most massive BHs.
Luminous quasars at these redshifts have a very
high large scale correlation length \cite{Shen07,Hennawi10,Shen10}, consistent with that
characterizing the spatial distribution of dark matter haloes as
massive as $\mhe \sim 10^{13}\, \msunhe$ \cite{Bonoli10},
in the assumption that quasar hosts are an unbiased tracer of
the underlying population of haloes of similar mass \cite{WL09}.
As first pointed out by \cite{White08}, this extremely high clustering amplitude,
combined with the corresponding space density, constrains the dispersion
in the luminosity-halo mass relation to be less than 50\% at 99\% confidence,
for the most conservative case of a 100\% duty cycle. In other words, all the haloes of mass
equal and above $\mhe \sim 10^{13}\, \msunhe$ at $z\gtrsim 4$ should be active, i.e., hosting very luminous quasars,
and the scatter in their luminosity-halo mass relation should be at least as small as the one in the local Universe
measured for the \mbh-\sis\ relation.

\cite{ShankarCrocce} further elaborated on the White et al. proposal,
showing that reproducing the observed
luminosity function and the high clustering of $z>3$ quasars,
also requires a high ratio between the luminosity in Eddington units $\lambda$ and the
radiative efficiency $\epsilon$. In other words, a radiative efficiency
of $\epsilon\gtrsim 0.2-0.3$ was favoured, for quasars radiating at a significant fraction
of their Eddington limit. Their method was based on
predicting the evolution of the BH mass
function directly from the evolution of the halo mass function,
once a given \mbh-\mh\ relation was specified.
The implied growth of BHs was then used to predict the
luminosity function through a continuity equation and an input mean
radiative efficiency and Eddington ratio.

While claims for a relatively strong evolution in the
BH-host galaxy mass relation are numerous, several studies performed by a variety
of groups with direct and indirect techniques consistently fail in detecting any parallel strong
evolution in the \mbh-\sis\ relation.
\cite{ShankarMsigma} adopted the local velocity dispersion function of spheroids,
together with their inferred age distributions,
to predict the velocity function at higher redshifts.
Taking the normalization of the \mbh-\sis\ relation to evolve as $(1+z)^\alpha$,
they computed the BH mass function associated with the velocity dispersion function
at each redshift, and compared to the cumulative BH mass density
inferred from the integrated quasar luminosity function (see \sect\ref{subsec|Accretion}).
This comparison, insensitive to the assumed duty cycle or Eddington ratio
of quasar activity, favoured a relatively mild redshift evolution, with $\alpha \sim 0.33$, with a positive
evolution as strong as $\alpha \gtrsim 1.3$ excluded at more than 99\% confidence level.

\begin{figure*}
    \center{\includegraphics[width=13truecm]{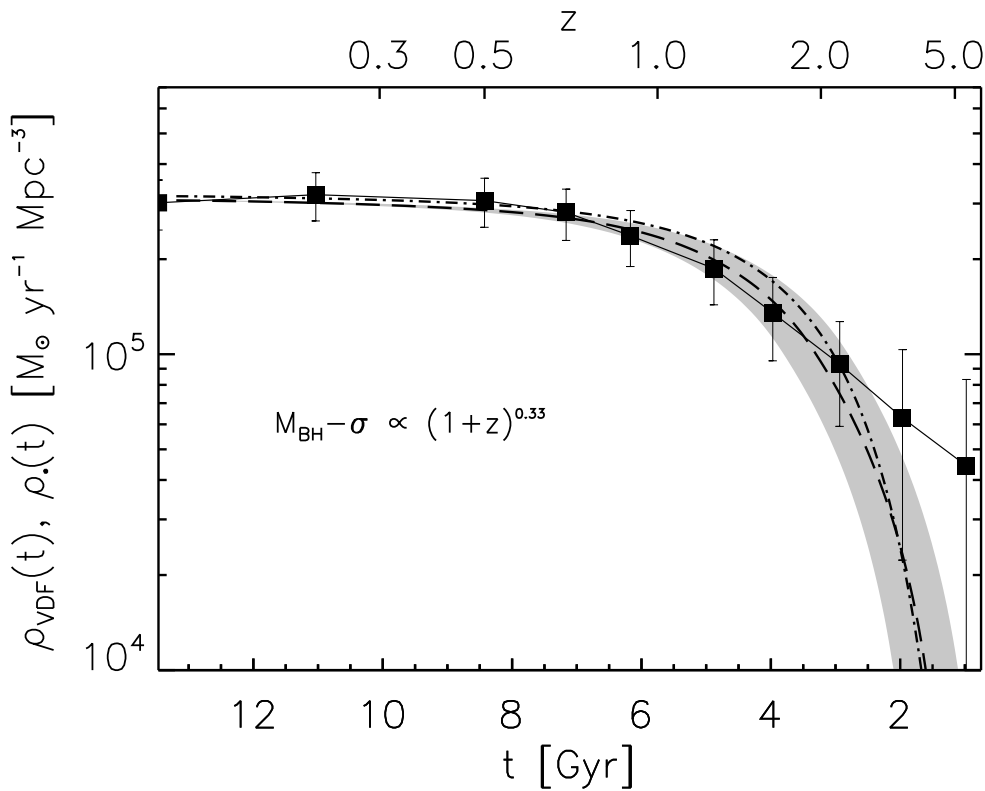}
    \caption{\emph{Solid} line with \emph{filled squares} is the BH accreted mass
  density at each redshift obtained from the convolution of the
  age-dependent, early-type velocity dispersion
  function, convolved with a redshift-dependent \mbh-\sis$\,$ relation, with
  a normalization scaling as $(1+z)^\alpha$.
  Long-dashed and dot-dashed lines are instead the BH mass densities
  inferred from integration
  of the AGN luminosity functions and a fixed radiative efficiency.
  The match in time between the two independent mass densities sets the constraint $\alpha \lesssim 0.3$, i.e.,
  nearly absent apparent evolution in the relation.}}
\label{fig|Shankar09Evol}
\end{figure*}

Their results are shown in \figu\ref{fig|Shankar09Evol}, where
the filled squares indicate the BH accreted mass
density at each redshift obtained from the convolution of the
age-dependent, early-type velocity dispersion
function, convolved with a redshift-dependent \mbh-\sis$\,$ relation, while
the long-dashed and dot-dashed lines are instead the BH mass densities
inferred from integration of the AGN luminosity functions and a fixed radiative efficiency.
We here note that the basic assumptions made by \cite{ShankarMsigma}
was that most of the stars in each nearby spheroid formed in a single episode and
the velocity dispersion \sis\ remained nearly constant at later epochs.
However, if the velocity dispersions of bulged galaxies increase
at higher redshifts paralleling their
apparently strong decrease in sizes \cite{Naab09,Shankar13},
the constraints on very low values of $\alpha$ would clearly become even tighter.
More recently, \cite{ZhangYuLu} performed a similar exercise, comparing accreted mass functions with
estimates of the local mass functions extrapolated to higher redshifts assuming some evolution
in the scaling relations. Their results yield a positive evolution for the correlation with stellar mass
consistent with \cite{Merloni10}, and null, or even negative results (anti-correlated with redshift)
for velocity dispersion, in full agreement with \cite{ShankarMsigma}.

Another piece of independent evidence in support of a null evolution in the
\mbh-\sis\ relation comes from direct spectral fitting in SDSS.
\cite{Salviander13} measured
BH masses and velocity dispersions from broad and narrow emission
lines in the quasar spectra of the SDSS Data Release 7, respectively,
finding minimal change in the relation for BHs in the
range $10^{7.5}< \mbhe/\msune < 10^9$ up to $z=1.2$.

The main conclusion of this section is that there seems to be growing
evidence for a significant positive evolution in the normalization of
the \mbh-\mstar\ relation, at least up to redshift $z\sim 2$, and some,
possibly non-linear and/or mass-dependent evolution in the scatter around it.
However, there is no apparent sign for any significant evolution in the normalization,
and possibly also scatter (otherwise we would have seen some evolution),
of the \mbh-\sis\ relation. How can this be possible?

\begin{figure*}
    \center{\includegraphics[width=15truecm]{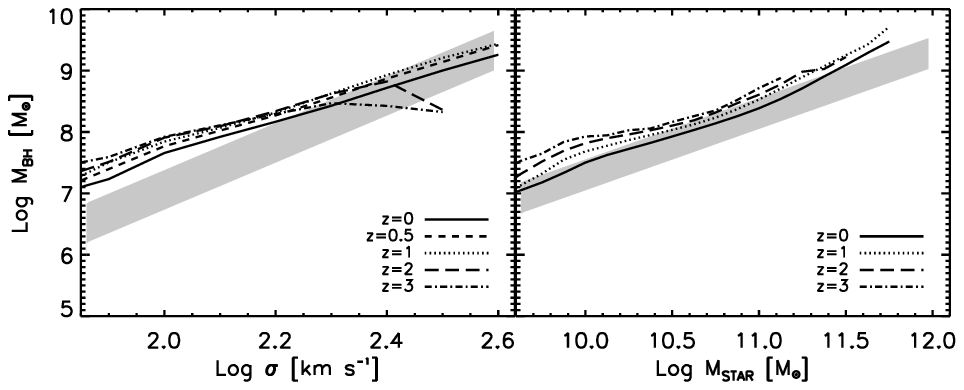}
    \caption{\emph{Left}: predicted \mbh-\sis\ relation
    at different redshifts, as labelled, for a model with a dark matter mass
    dependent \sis\ for galaxies with bulge-total ratios $B/T>0.9$.
    The \emph{grey stripe} indicates the fit by \cite{Tundo07}
    with its intrinsic scatter. \emph{Right}: predicted \mbh-\mstar\ relation
    at the same redshifts for galaxies with $B/T>0.9$.
    The \emph{grey stripe} shows a linear relation of the type $\mbhe=2\times 10^{-3} \mstare$
    with some scatter, indicative of what suggested by a variety of local data.}}
\label{fig|Shankar13}
\end{figure*}

In order to properly consider the problem, one should insert the issue of
redshift evolution in BH scaling relations
within the broader context of structural evolution of massive galaxies.
It is now well established that early-type galaxies show a strong half-light
radius evolution when scaling up with redshift, becoming progressively more
compact by a factor of a few at redshift $z\gtrsim 1$ with respect to their
local counterparts of similar stellar mass \cite{Cimatti12,Huertas13a}.
A reduction in size at fixed stellar mass should be paralleled by some
increase in the velocity dispersion. Thus, at fixed BH mass,
one would expect, if anything, a
\emph{decrease} in the normalization of the \mbh-\sis\ relation, more or less
strong depending on how much mass is actually accreted onto the BH during the gas-rich,
high-redshift phase of massive galaxy formation \cite{Granato04,Malbon07}.
However, if BHs are effectively mapped to lower stellar masses naturally characterized
by lower velocity dispersion, this would conspire to erase or even reverse the
predicted evolution in the observed \mbh-\sis\ relation, thus reconciling the
separate observations on the disparate degree of evolutions on
BH mass, and velocity dispersion/size of the host galaxy.
This line of thought has been recently more quantitatively confirmed by
\cite{Shankar13}, who self-consistently computed
sizes and velocity dispersions within the \cite{Guo11} semi-analytic model,
finding a positive evolution in the \mbh-\mstar\ relation, and milder, but still positive,
in the \mbh-\sis\ relation (see \figu\ref{fig|Shankar13}).

\section{How to use Black Hole Demography to constrain Models}
\label{sec|Models}

\subsection{Semi-empirical, continuity equation models}
\label{subsec|Accretion}

Semi-empirical models, although clearly
more limited in scope than more advanced galaxy formation models,
can still provide useful physical insights on galaxy-BH evolution,
which can in turn be further interpreted within more advanced galaxy evolution models.
To this purpose, the original proposal by \cite{Soltan,Salucci99} of matching
the local and accreted BH mass densities to limit the average radiative efficiency of BHs,
has been recently developed by \cite{Shankar13acc} into a more comprehensive
semi-empirical model.

The demography of the BH population through
time, is numerically computed by self-consistently
solving the following continuity equation \cite{Cavaliere71},
\begin{equation}
\frac{\partial n_{\rm BH}}{\partial
t}(M_{\rm BH},t)=-\frac{\partial (\langle \dot{M}_{\rm BH}\rangle
n_{\rm BH}(M_{\rm BH},t))}{\partial M_{\rm BH}}\, .
    \label{eq|conteq}
\end{equation}
Here $\langle \dot{M}_{\rm BH}\rangle$ is the mean accretion
rate (averaged over the active and inactive populations) of the
BHs of mass \mbh\ at time $t$ \cite{SmallBlandford}.

\cite{Shankar13acc} generalized these continuity-equation models to allow for any input
mass and/or redshift-dependent
radiative efficiency, and observationally motivated
Eddington ratio distributions $P(\lambda|\mbhe,z)$, with
$\lambda \equiv L/L_{\rm Edd}$ defined as the ratio between bolometric and Eddington luminosity.
Through this advanced semi-empirical approach,
these Authors found that reproducing the high observed fractions of active galaxies
at low redshift requires a characteristic Eddington ratio
that steadily declines at late times, in a possible mass-dependent manner (more massive BHs
having lower $\lambda$ than less massive counterparts at similar epochs).
In other words, at fixed mass, BHs become progressively less efficient in time in
shining at high luminosities, either because the triggering mechanisms become rarer \cite{Shen09},
and/or they change with time \cite{DraperBallantyne12},
and/or simply because the fuelling rate continuously drops \cite{Vittorini05,Cav07}.

\begin{figure*}
    \center{\includegraphics[width=15truecm]{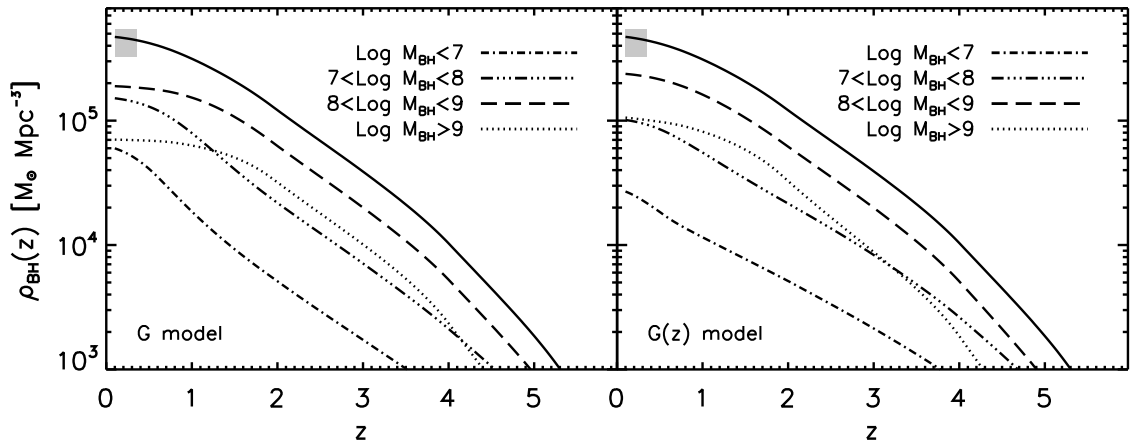}
    \caption{Cumulative BH mass density as predicted from the continuity equation models
    of \cite{Shankar13}. \emph{Left}: results for a model with input Gaussian Eddington ratio distribution
    and constant with time. \emph{Right}: same Gaussian model as in the left panel,
    with peak $\lambda$ steadily decreasing with cosmic time.
    The \emph{solid} lines represent the total cumulative BH mass density
    as a function of redshift, while the other lines
    mark the contributions from BHs in different mass bins, as labelled.
    The \emph{grey bar} indicates the values and systematic
	uncertainties in the total local mass density in black holes estimated
	by \cite{SWM}. When the characteristic $\lambda$ decreases the low
    mass BHs accrete less mass, while the more massive accrete more,
    especially at $z \lesssim 1$.}}
\label{fig|GzModel}
\end{figure*}

\figu\ref{fig|GzModel} presents the cumulative BH mass density as a function of redshift
for two continuity equation models taken from \cite{Shankar13acc}.
The models share same input radiative efficiency and Gaussian Eddington ratio distributions,
but differ in having the characteristic $\lambda$ (i.e., the median of the Gaussian),
in one case constant at $1/3$ (left panel), and steadily decreasing with cosmic time
(from Eddington to strongly sub-Eddington) in the other case (right panel).
The solid lines show the cumulative total BH mass density as a function of redshift,
while the other lines indicate the mass
density accreted in selected bins of current BH mass, as labelled.
The main relevant feature arising from the comparison between the two panels
is that in the former case (constant $\lambda$), BHs more massive
than $\mbhe \gtrsim 10^8\, \msune$ stop accreting below $z\lesssim 1$,
while in the latter (decreasing $\lambda$) they continue to significantly
grow in mass, at the expense of the less massive ones which grow much less.
The solid grey square indicates the systematic uncertainties
in the total local BH mass density estimated by \cite{SWM}.

\figu\ref{fig|BHmassFunctionPseudo} reports the predictions of the $z=0$ BH mass
function predicted by the two above mentioned accretion models,
compared with the empirical estimates discussed in \sect\ref{sec|scalingrelations}.
The no evolution-model (long-dashed line) well matches the local BH mass function
inferred on the assumptions that all galaxies follow similar scaling relations (solid, red lines).
The model with decreasing characteristic $\lambda$ produces instead significantly
less numerous BHs at the low mass end. Decreasing $\lambda \propto L/\mbhe$ in fact,
maps a given luminosity to more massive and rarer BHs, thus progressively
limiting the growth of less massive BHs boosting the accretion onto the most massive ones.
As extensively discussed and reviewed by \cite{Shankar13acc}, a decreasing
Eddington ratio is currently favoured by several direct observational signatures,
such as the Eddington ratio distributions of local SDSS active galaxies \cite{Kauffmann09},
and the high fraction of massive active galaxies \cite{Goulding09}.

\subsection{More complex evolutionary models}
\label{subsec|MoreModel}

Besides accretion models, a variety of more or less complex models
for the evolution of the BH population has been put forward in the Literature
along the years. Most of them are based on triggering mechanisms associated to
mergers and/or flybys \cite{Marulli08,Vittorini05,Menci06,Malbon07,ShenMergers,Shankar10,Neistein13},
as well as in-situ processes, such as more
or less strong disc instabilities \cite{Malbon07,Bournaud11b}, or other processes \cite{Kollmeier06,Lapi06}.
Models to probe BH cosmological evolution via gas accretion and mergers
can be highly sophisticated, especially if BHs are modelled within the
already complex, and still not fully understood, net of host galaxies and dark matter haloes
\cite{Fontanot11AGN,Fanidakis13}. Due to the diverse physical assumptions
and the non-trivial degeneracies induced by the large set of
underlying parameters characterizing and shaping galaxy formation models, it is still very
hard to constrain the successful physical models of galaxy and BH formation.

To mention some of the most recent results in the topic, \cite{Hirsch12}
within the context of a full semi-analytic model for galaxy formation,
suggest a scenario in which disc instabilities are the main
driver for moderately luminous Seyfert galaxies at low redshift, while major mergers
are the main trigger for luminous active galaxies. Similar conclusions
were reached by \cite{DraperBallantyne12} in the context of a semi-empirical
model, based on combining an observationally motivated AGN triggering rate
and a theoretical AGN light curve. They found major mergers to be insufficient
to account for the entire AGN population, and claimed non-merger processes,
such as secular mechanisms, to be the dominant AGN triggering mechanism at $z \lesssim 1-1.5$.

While the basic notion that major mergers may not explain the full
AGN demography seems to be confirmed by many independent works
\cite{ShenMergers,Shankar10,Treister12},
minor mergers have been found to still represent a rather successful mechanism to
reproduce the full AGN luminosity function and clustering properties
\cite{Marulli08,Bonoli09,Bonoli10}, as well as their connection
to star formation rates \cite{Neistein13}.

\cite{ShankarPseudo} have analyzed the predictions of two state-of-the-art hierarchical galaxy
formation models. In the first one BHs grow only via major and minor mergers \cite{Guo11},
while in second one \cite{Bower06} BHs are allowed to grow also via disc instabilities.
Their study highlighted the fact the model in which BHs always closely follow the growth of
their host bulges, also during late disc instabilities (i.e., bars),
produces too narrow a distribution of BHs at fixed stellar mass to account for the numerous
low-mass BHs now detected in later-type galaxies (see \sect\ref{sec|scalingrelations}).
Models with a looser connection between BH growth and bar instability
instead predict the existence of a larger number of under-massive BHs, in better agreement with the observations.
Simulations and direct observations support the presence and growth of
stellar bars in gas-poorer systems \cite{Atha13},
thus possibly disfavouring a strong link between BH gas fuelling and stellar bar growth.
On the other hand, clumpy accretion of gas clumps towards the centre of gas-rich and turbulent,
high redshift discs, could still represent a viable mechanism to grow classical bulges and their central BHs
\cite{DiMatteo08,Bournaud11b}.

It is thus clear that secure knowledge of the
the slopes and intrinsic scatters of BH scaling relations can impose valuable constraints to alternative models
of BH evolution. For example, the exact slope in the \mbh-\sis\ relation could
allow to discern between a radiative- against a momentum-driven AGN feedback \cite{KingPringle}.
If instead repeated BH dry mergers are the primary cause
behind the growth of the most massive BHs,
then BH scaling relations should progressively tighten with increasing
mass \cite{Peng07,JahnkeMaccio} (but see also
\cite{Ciras05}). However, the present sparse sample
of local BHs offers only tentative evidence for this \cite{McConn13}.

Additional, independent constraints on BH models can be derived from AGN clustering.
On empirical grounds, as reviewed by \cite{ShankarReview},
sharp AGN clustering measurements offer
a unique constraint to the duty cycles of active galaxies and scatters around the median BH
scaling relations \cite{HH01,MW01}.
As anticipated in \sect\ref{sec|EvolScalingRelations},
reproducing the strong observed clustering of $z=4$ quasars, for example,
requires duty cycles close to unity and minimal scatter between luminosity and
halo mass. On the other hand, reconciling the lower values of the correlation
length of luminous quasars at $z\approx 1.5$, requires significant scatter
between luminosity and halo mass to lower the predicted clustering amplitude
\cite{Shankar10shen,ConroyWhite}.

Merger models can broadly reproduce the clustering of quasars at nearly all epochs and scales
\cite{Bonoli09,ShenMergers}.
This is because mergers are most efficient in haloes of masses around
$\mhe \sim 3 \times 10^{12}\, \msune$, which is the typical mass scale inferred from direct clustering
measurements of quasars of nearly all luminosities and redshifts \cite{Shen09}.
However, in the last years there has been mounting evidence for
the clustering of X-ray AGN \cite{Cappelluti12},
to be significantly higher than the corresponding values of optically
selected AGN at the same redshifts \cite{Allevato11,Allevato12,Kou13},
consistent with dark matter halo masses up to one order of magnitude higher than those typical of optical quasars.
While clearly larger samples are needed to better control eventual systematic observational biases,
such as cosmic variance \cite{Gilli05}, it is still worth exploring some
possible physical causes behind these apparent discrepancies.
\cite{Fanidakis13} claim that an additional channel
for BH accretion is required, namely hot-halo mode,
which is disassociated to the cold accretion during disc instabilities and galaxy mergers discussed above.
In their model the hot-halo mode becomes prominent in dark matter haloes
with masses greater than $\mhe \gtrsim 10^{12.5}\, \msune$,
giving rise to a distinct class of moderate luminosity AGN that inhabit rich clusters and superclusters.

\section{Conclusions}
\label{sec|conclu}

In this contributed paper I have reviewed several key topics on the demography of BHs.
I started by reviewing the latest results on the local scaling relations between BHs and their
host galaxies, concluding that there is still significant mismatch among the results of different groups,
mainly because of the still limited sample available. This in turn poses challenges to determine a complete
and secure calibration of the local BH mass function which still presents a systematic uncertainty in the
normalization and shape. The latter is especially true at the low mass end, where the role of BHs in later-type
galaxies and/or pseudobulges is still not properly understood.
I then continued reviewing the continuous cumulative evidence for
an evolving \mbh-\mstar\ relation at high redshifts, and a non-evolving \mbh-\sis\ relation,
which would be consistent with the coupled strong evolution of the size function of massive spheroids.
Continuity accretion models can account for the full local BH mass function, and favour steadily decreasing
Eddington ratios with cosmic time. More complex models based on mergers and/or strong high redshift disc
instabilities are consistent with the local demography, while models with bar instabilities tend to be disfavoured.

\section*{Acknowledgments}
My warmest thanks to unique collaborators on these topics: Viola Allevato, Mariangela Bernardi,
Laura Ferrarese, Alexis Finoguenov, Zoltan Haiman, Ronald L\"{a}sker, Smita Mathur,
Jordi Miralda-Escud\'{e}, Jorge Moreno, and David Weinberg.
I also wish to thank Alister Graham, Andrea Lapi, and Youjun Lu for many interesting and insightful discussions.
I acknowledge support from a Marie Curie Fellowship.

\section*{References}
\bibliographystyle{unsrt}
\bibliography{../../RefMajor_Rossella}

\end{document}